\documentclass{article}

    \usepackage[final]{tackling_climate_workshop_style}

\usepackage[utf8]{inputenc} 
\usepackage[T1]{fontenc}    
\usepackage{hyperref}       
\usepackage{url}            
\usepackage{booktabs}       
\usepackage{amsfonts}       
\usepackage{nicefrac}       
\usepackage{microtype}      

\usepackage{graphicx}
\usepackage{amsmath}
\usepackage{tabularx}

\title{Interactive Atmospheric Composition Emulation for Next-Generation Earth System Models}

\author{%
    Mohammad H. Erfani\thanks{Corresponding author (\texttt{se2639@columbia.edu})} \\
    Center for Climate Systems Research\\
    Columbia University\\
    New York, NY 10025 \\
    \And
    Kara D. Lamb \\
    Department of Earth and Environmental Engineering \\
    Columbia University \\
    New York, NY 10027\\
    \And
    Susanne E. Bauer \\
    NASA Goddard Institute for Space Studies\\
    New York, NY 10025 \\
    \And
    Kostas Tsigaridis \\
    Center for Climate Systems Research\\
    Columbia University\\
    New York, NY 10025 \\
    \And
    Marcus van Lier-Walqui\\
    Center for Climate Systems Research\\
    Columbia University\\
    New York, NY 10025\\
    \And
    Gavin Schmidt\\
    NASA Goddard Institute for Space Studies\\
    New York, NY 10025 \\
}

\begin{document}

\maketitle

\begin{abstract}
    Interactive composition simulations in Earth System Models (ESMs) are computationally expensive as they transport numerous gaseous and aerosol tracers at each timestep. This limits higher-resolution transient climate simulations with current computational resources. ESMs like NASA GISS-ModelE3 (ModelE) often use pre-computed monthly-averaged atmospheric composition concentrations (Non-Interactive Tracers or NINT) to reduce computational costs. While NINT significantly cuts computations, it fails to capture real-time feedback between aerosols and other climate processes by relying on pre-calculated fields. We extended the ModelE NINT version using machine learning (ML) to create Smart NINT, which emulates interactive emissions. Smart NINT interactively calculates concentrations using ML with surface emissions and meteorological data as inputs, avoiding full physics parameterizations. Our approach utilizes a spatiotemporal architecture that possesses a well-matched inductive bias to effectively capture the spatial and temporal dependencies in tracer evolution. Input data processed through the first 20 vertical levels (from the surface up to 656 hPa) using the ModelE OMA scheme. This vertical range covers nearly the entire BCB concentration distribution in the troposphere where significant variation on short time horizons due to surface level emissions is observed. Our evaluation shows excellent model performance with R² values of 0.92 and Pearson r of 0.96 at the first pressure level. This high performance continues through level 15 (808.5 hPa), then gradually decreases as BCB concentrations drop significantly. The model maintains acceptable performance even when tested on data from entirely different periods outside the training domain, which is a crucial capability for climate modeling applications requiring reliable long-term projections. These results confirm that our approach successfully shifts the paradigm from simple numerical solver mimicry to spatio-temporal modeling, offering significant improvements in forecasting capability for long-term climate simulations.
\end{abstract}

\section{Introduction}
    Earth System Models (ESM) are foundational tools in climate science, combining numerical representations of atmospheric fluid dynamics and other Earth physical processes, such as the ocean, land surface, and cryosphere, on three-dimensional grids~\cite{erfani2024spatiotemporal}. Despite their importance, ESMs still have significant shortcomings, including imperfect physical parameterizations, insufficient resolution to resolve fine-scale weather phenomena, and prohibitively high computational costs for operational forecasting. To address these limitations, researchers have increasingly turned to advanced machine learning (ML) techniques. These efforts were pioneered by~\cite{weyn2019can,weyn2020improving,weyn2021sub}, building upon foundational studies such as~\cite{dueben2018challenges,scher2018toward,scher2018predicting,scher2019weather}. Deep Learning Weather Prediction (DLWP) models utilize deep convolutional neural networks (CNNs) for globally gridded weather prediction. The focus is on developing a data-driven ML-based model that can be iteratively stepped forward, similar to traditional numerical weather prediction (NWP) models, to simulate atmospheric states at arbitrarily long lead times. 

    Recent studies has continued to follow traditional numerical solver approaches while working to better represent global grid data. These efforts have explored various representations including equiangular gnomonic cubed spheres~\cite{weyn2020improving}, spherical harmonic functions~\cite{pmlr-v202-bonev23a}, and multi-mesh graph structures~\cite{lam2023learning}. Additionally, these studies have enhanced model performance in the spatial domain by incorporating advanced machine learning components that incorporate physical intuition like Fourier neural operators~\cite{li2020fourier, pathak2022fourcastnet}, graph neural networks (GNNs)~\cite{keisler2022forecasting}, and vision transformers~\cite{benson2025atmospheric}. The primary aim of these approaches is to better capture spatial relationships within 3D grids by using ML modules that provide stronger inductive biases for understanding connections between grid points. However, these methods still rely on ``autoregressive rollout'' approaches similar to traditional weather prediction systems. This means they generate long-term weather forecasts by repeatedly feeding their own predictions back as input to create extended sequences of weather states. Almost all studies use this mechanism to handle time-dependent aspects of weather prediction. Unfortunately, this approach is clearly inadequate and represents a major limitation for stable long-term forecasting. This is particularly problematic for atmospheric dynamics, which are fundamentally time-dependent processes. The absence of architectural components specifically designed to capture trends, seasonal patterns, and cyclical behaviors significantly restricts the model's ability to make accurate forecasts over extended time periods.
     

    In the spatial domain, most studies focus their design efforts primarily on horizontal interactions, treating vertical structure as generic feature channels. The models typically receive an array of input values (prognostic/diagnostic variables) per horizontal grid cell, which they process in a latent space, presumably to account for column physics and vertical mixing. GraphCast~\cite{lam2023learning} stands as an exception, introducing a systematic approach for communicating information between different vertical layers through the message-passing mechanism of GNNs. The choice of appropriate ML components (modules) that can capture the inherent characteristics of a specific task is crucial for model performance. Consider how attention mechanisms~\cite{vaswani2017attention} revolutionized natural language processing by enabling models to dynamically focus on relevant parts of input sequences regardless of distance, outperforming previous sequential models like bidirectional LSTMs that processed sentences in a rigid, linear fashion~\cite{peters2018deep}. Similarly, developing specialized vertical modeling components could unlock significant breakthroughs in atmospheric modeling by properly accounting for column physics, i.e., capturing the complex multi-level interactions and dependencies that govern atmospheric dynamics. Just as attention mechanisms provided NLP models with the ability to handle long-range dependencies in text, purpose-built ML components for vertical atmospheric modeling could better represent the unique physical processes occurring across different atmospheric layers.

    We propose spatiotemporal ML as an alternative approach to autoregressive rollout methods to simultaneously capture the inherent high spatial and temporal dependencies in climate data. The temporal module of spatiotemporal ML models, regardless of type (recurrent operations~\cite{elman1990finding}, long short-term memory~\cite{graves2012long}, temporal CNNs~\cite{bai2018empirical}, or attention~\cite{vaswani2017attention}), generally incorporates an appropriate mechanism for capturing temporal dependencies. To address the gap in explicit architectural components for vertical atmospheric coupling, we introduce a preprocessing module that processes two different modalities (2D emission and associated 3D forcings) and fuses them into an identical dimension~\cite{tan2023temporal, wang2018recovering}. The output can be fed into any spatiotemporal model architecture, where ConvLSTM~\cite{shi2015convolutional} is adopted in this study. This study focuses on carbonaceous aerosols, particularly Black Carbon from biomass burning (BCB), the most significant absorbing aerosol species in terms of its direct radiative forcing impacts.
    
\section{Methodology}
    In this work, we emulate the prediction of BCB (Black Carbon from Biomass Burning) transport for the first 20 vertical levels (up to 600 hPa) of the NASA GISS-E3 ESM model (tropospheric portion). Specifically, we aim to forecast the 3D field of BCB concentration in the atmosphere at timestep $t$, given meteorological inputs (e.g., wind components and precipitation) and BCB surface fluxes from the previous 47 timesteps. The list of candidate variables, their description, units and dimensions on spatial domain are showed in Appendix~\ref{sec:appendix-1}.
    
    \subsection{Dataset Description}
        To train the ML model, the NASA GISS-E3 (ModelE) was run with prescribed sea surface temperature (SST) and sea ice fraction for three different time periods. First, the pre-industrial era from 1851 to 1853 (3 years), yielding 3 years of data (E3OMA1850M). For this period, monthly-averaged BCB concentrations were interpolated to daily intervals, as no daily inventory exists for this time period. The second dataset spans 2010 to 2012 (3 years) (E3OMA2010D). Since this period falls within the satellite era, daily BCB concentrations were used for this run. The third dataset covers 2020 and 2021 (2 years) (E3OMA2020D). Like the 2010 dataset, this run is based on daily BCB concentrations and is considered for testing the model on entirely different time periods outside the training domain.

    \subsection{Model Architecture}
        We propose a spatiotemporal ML approach to project the effects of interactive emissions on climate forcing. Before data is fed into the spatiotemporal model, they are passed into a preprocessing module for spatial feature extraction and modality fusion. To do this, we employ spatial encoders~\cite{tan2023temporal} that process two data modalities (2D surface variables and 3D forcing) individually and transform them into an identical dimension in the latent space. Results are then fused via Spatial Feature Transform (SFT)~\cite{wang2018recovering} which generally modulates (affine transform) the 3D forcing given 2D variables. Since the horizontal dimensions are preserved throughout this processing pipeline, the output maintains four dimensions (i.e., time, features, latitude, and longitude) enabling seamless integration with any spatiotemporal model architecture regardless of their spatial domain representation approach~\cite{tan2023openstl}. Here, we employ ConvLSTM~\cite{shi2015convolutional} with a rectangular latitude-longitude grid representation, as extensive prior research has established effective approaches in this area and spatial representation was not our primary focus. The details of the model architecture is elaborated in the Appendix~\ref{sec:appendix-2}.
        
\section{Results}
    Table~\ref{tab:model_comparison} presents the performance comparison of the proposed model in terms of R-squared ($R^2$) across different test periods and pressure levels. The $R^2$ values are computed at each timestep for all 20 pressure levels during the test period by comparing ground truth and predicted 2D fields. The values reported in the table represent the temporal average of $R^2$ across all timesteps for each pressure level. Across all test years, the model performance consistently decreases at higher pressure levels (lower atmospheric pressure). We observe a strong correlation between BCB concentration magnitude and $R^2$ values at each level (e.g., correlation coefficient of 0.92 for the 2012 test period). Two primary factors contribute to this performance degradation:

    \begin{table}[htbp]
        \centering
        \caption{Model performance ($R^2$) comparison across different test years and pressure levels.}
        \label{tab:model_comparison}
        \begin{tabularx}{\textwidth}{c|*{12}{X}}
            \hline
            \textbf{Year} & \textbf{979} & \textbf{969} & \textbf{959} & \textbf{949} & \textbf{939} & \textbf{888.5} & \textbf{808.5} & \textbf{784} & \textbf{756.5} & \textbf{726} & \textbf{692.5} & \textbf{656} \\ 
            \hline
            \textbf{1853} & 0.92 & 0.90 & 0.87 & 0.86 & 0.85 & 0.84 & 0.79 & 0.76 & 0.73 & 0.68 & 0.58 & 0.37 \\
            \textbf{2012} & 0.88 & 0.85 & 0.81 & 0.79 & 0.78 & 0.76 & 0.70 & 0.68 & 0.65 & 0.61 & 0.55 & 0.39 \\
            \textbf{2020} & 0.81 & 0.77 & 0.73 & 0.70 & 0.69 & 0.67 & 0.61 & 0.59 & 0.57 & 0.52 & 0.43 & 0.24 \\
            \textbf{2021} & 0.82 & 0.80 & 0.76 & 0.74 & 0.73 & 0.70 & 0.63 & 0.61 & 0.58 & 0.55 & 0.50 & 0.36 \\ 
            \hline
        \end{tabularx}
    \end{table}


    i) During training, the target variable is normalized using BCB concentration statistics from the first pressure level. This approach is necessary because BCB concentrations naturally decrease with altitude across atmospheric levels. Using level-specific normalization would artificially inflate concentration values at higher altitudes, creating misleading representations of the atmospheric profile. However, this global normalization strategy introduces a systematic bias favoring lower atmospheric levels, where concentrations are naturally higher. 

    ii) The BCB concentration differences between the first and last pressure levels span an order of magnitude. At higher atmospheric levels, smaller absolute concentrations amplify relative prediction errors, significantly impacting $R^2$ values. This makes the model appear less accurate at higher levels even when absolute errors remain comparable. To verify this effect, we examined the Pearson correlation coefficient ($r$), which is less sensitive to scale differences. For the 2012 test period, the difference between Pearson $r$ values at the first level (0.94) and the highest level (0.73) is only 0.21, much smaller than the corresponding $R^2$ gap. This indicates the model maintains strong predictive capability throughout the vertical profile, despite $R^2$ disproportionately penalizing predictions at levels with lower absolute concentrations.
    
\begin{figure}[htbp]
        \centering
        \includegraphics[width=1\linewidth]{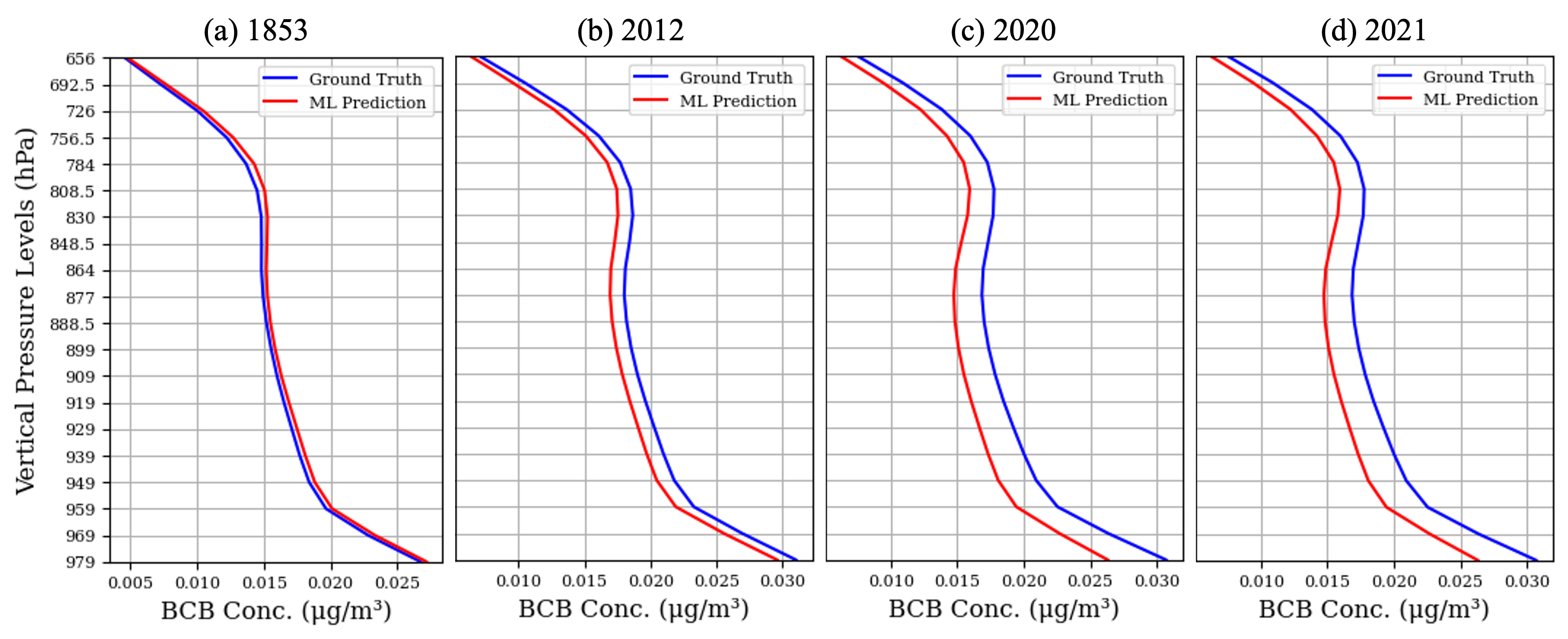}
        \caption{Vertical profiles of BCB concentration across 20 atmospheric levels for four different years: (a) 1853, (b) 2012, (c) 2020, and (d) 2021.}
        \label{fig:vertical_profile}
    \end{figure}

    This discrepancy becomes more pronounced in Figure~\ref{fig:vertical_profile}, which compares BCB concentration profiles across 20 vertical levels for ML prediction versus ground truth data.
    The profiles demonstrate the model ability to capture the general vertical distribution pattern of BCB across different time periods. In 1853, model predictions closely match observations throughout the vertical column. In contrast, for more recent years (2012-2021), the model consistently underestimates BCB concentrations, particularly at lower and middle atmospheric levels, while showing improved agreement at higher levels. 
    
    \subsection{Annual Mean BCB Column Burden}
\begin{figure}[htbp]
        \centering
        \includegraphics[width=1\linewidth]{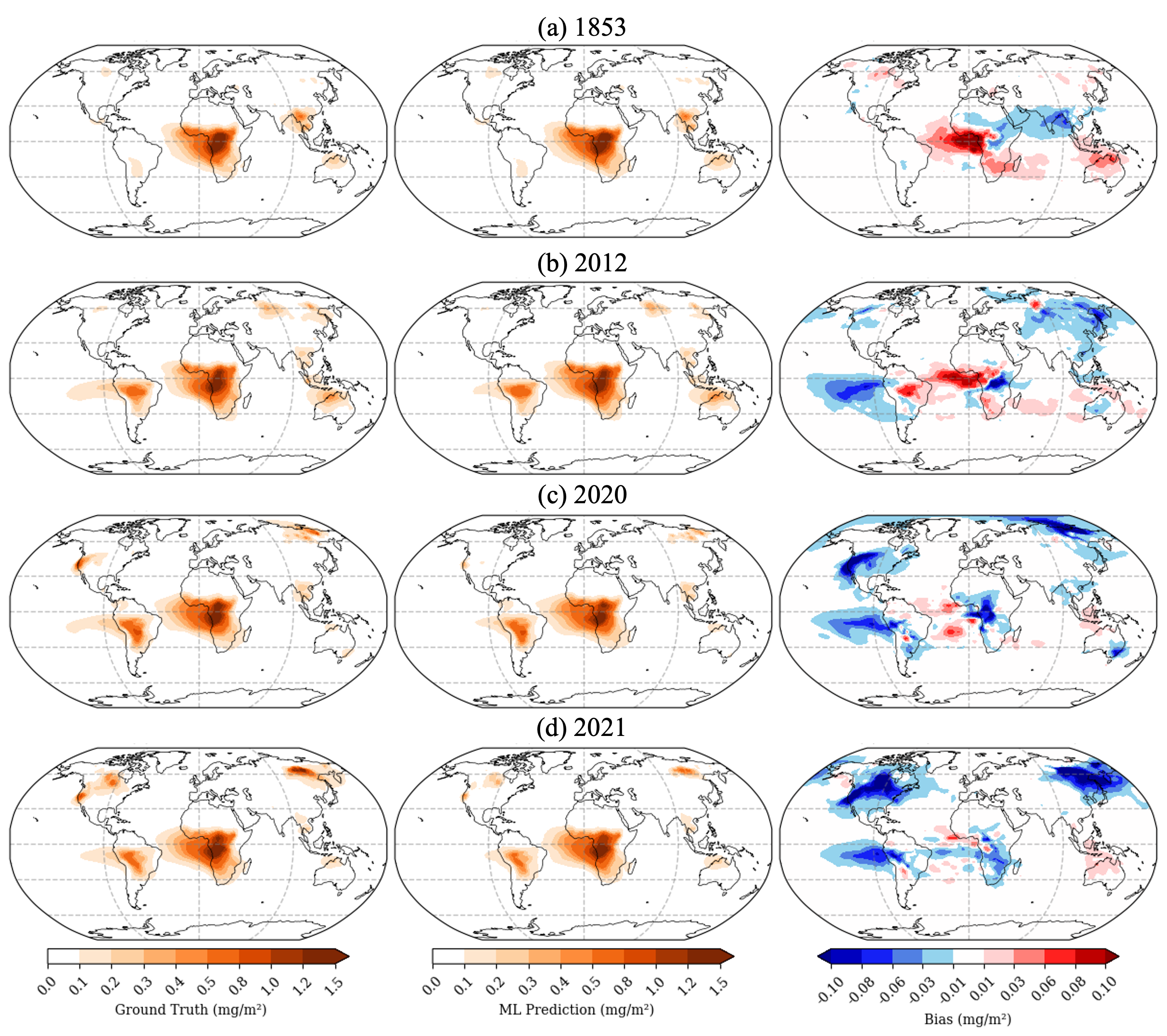}
        \caption{Annual mean BCB column burden. The first column is the ground truth from ModelE simulations. The second column is the machine learning predictions, and the third column represents the bias (prediction - ground truth).}
        \label{fig:column_burden}
    \end{figure}

    To quantify the overall performance of the model across different regions and time periods, we compare the column burden of ground truth (ModelE simulation outputs) with machine learning (ML) predictions (see Figure~\ref{fig:column_burden}). In general, the ML-predicted column densities tend to be biased low compared to ground truth, with this underestimation becoming more pronounced over time. This trend, however, is less evident in tropical regions with high biomass burning activity, such as the Amazon and Central/Southern Africa.

    The model struggles particularly in regions with newly emerging or episodic activity during the test period, such as the US and northeast Russia. Specifically, in 2020 and 2021, the model substantially underestimates values in the US West, a period marked by unprecedented wildfire events that emitted record-breaking amounts of black carbon aerosols and particulates. For instance, the 2020 US wildfire season saw five of the six largest fires on record in California, as well as massive fires in Oregon and Washington, with total burned areas exceeding 8 million acres and resulting in significant spikes in atmospheric black carbon. Similarly, 2021 recorded exceptionally high levels of wildfire carbon emissions, particularly from boreal and Western US fires, driving steep increases in atmospheric soot content~\cite{sannigrahi2022examining}. For northeast Russia, the historical context remains important. According to the statistics of the Global Wildfire Information System (GWIS), the summer of 2012 was the most severe wildfire season in a decade, with extensive burning across the Siberian taiga fueled by abnormal heat and drought conditions, causing dramatic increases in regional black carbon levels. 

\section{Conclusion}
    In this study, we investigated a spatiotemporal machine learning approach as an alternative to autoregressive rollout methods for computing atmospheric composition dynamics, with particular emphasis on biomass-burning black carbon (BCB). Unlike autoregressive rollout approaches, which lack the necessary architectural components to capture long-term time-dependent behavior, our method directly addresses the underlying physical problem through end-to-end machine learning. Our approach incorporates specialized architectural modules that encode inductive biases aligned with the inherent `spatial' and `temporal' dependencies of atmospheric transport and chemistry. By embedding these physical constraints directly into the model architecture, we achieve predictions that remain physically realistic over extended time horizons, which is a critical requirement for climate modeling applications where simulations must maintain reliability across multi-decadal to centennial projections.
    
    Our findings confirm our central hypothesis that "architectural provisions" for temporal evolution and vertical atmospheric coupling produce corresponding improvements in model accuracy and physical consistency. These gains demonstrate that architectural design for time and height is crucial, complementing the necessary efforts to respect the Earth's spherical geometry through appropriate projections, multi-mesh graph structures, or the use of spherical harmonic transforms.

\bibliographystyle{unsrt}
\bibliography{ref.bib}

\newpage \appendix
\section{Dataset Description} \label{sec:appendix-1}
    The atmospheric component of ModelE is configured for this study with a horizontal resolution of 2$^\circ$ latitude by 2.5$^\circ$ longitude and 62 vertical layers. The model's temporal resolution is set to half-hour intervals. Model outputs comprise prognostic and diagnostic variables required to consider all sub-processes in the BCB transport model, including horizontal advection, vertical convection, and sink/source terms. The list of variables is shown in Table~\ref{tab:model_variables}.

    \begin{table}[htbp]
        \centering
        \caption{ModelE variables used in BCB transport modeling}
        \label{tab:model_variables}
        \begin{tabular}{llll}
            \hline
            Variable Name       & Description                       & Unit                               & Dim \\ \hline
            axyp                & Gridcell area                     & $\text{m}^2$                       & 2D  \\
            landfr              & Land fraction                     & \%                                 & 2D  \\
            prsurf              & Surface pressure                  & hPa (mb)                           & 2D  \\
            pblht$_{\text{bp}}$ & Planetary boundary layer height   & m                                  & 2D  \\
            shflx               & Sensible heat flux                & $\text{W}/\text{m}^2$              & 2D  \\
            lhflx               & Latent heat flux                  & $\text{W}/\text{m}^2$              & 2D  \\
            BCB$_{\text{src}}$  & BCB source                        & $10^{-12}$ kg m$^{-2}$ s$^{-1}$    & 2D  \\
            u                   & East-west velocity                & m/s                                & 3D  \\
            v                   & North-south velocity              & m/s                                & 3D  \\
            omega               & Pressure vertical velocity        & hPa/s                              & 3D  \\
            p$_{\text{3D}}$     & Pressure on model levels          & mb                                 & 3D  \\
            z                   & Layer altitude (above MSL)        & m                                  & 3D  \\
            t                   & Temperature                       & K                                  & 3D  \\
            th                  & Potential temperature             & K                                  & 3D  \\
            q                   & Specific humidity                 & kg/kg                              & 3D  \\
            prec$_{\text{3D}}$  & Precipitation                     & mm/d                               & 3D  \\
            cfrad               & Cloud fraction                    & \%                                 & 3D  \\
            mcuflx              & Convective updraft mass flux      & kg m$^{-2}$ s$^{-1}$               & 3D  \\
            airmass             & Air mass density                  & kg m$^{-2}$                        & 3D  \\
            BCB                 & BCB mixing ratio                  & $10^{-11}$ kg/kg$_{\text{air}}$    & 3D  \\ \hline
        \end{tabular}
    \end{table}
    
    According to Equation~\ref{eq:transport_model} and depending on how the convection term is parameterized, the list of required variables may vary. Several of the aforementioned variables were excluded based on preliminary analysis, including estimation of mutual information for continuous target variables, univariate linear regression tests, recursive feature elimination, and feature importance analysis using Random Forest. This preliminary analysis was necessary because loading all the mentioned 2D and 3D variables creates a significant bottleneck in the training process.
    
    Eventually, we selected 2 static variables representing the area of each gridcell and the fraction of land, sea, and ice; four 2D variables existing at the surface including planetary boundary layer height, sensible heat flux, latent heat flux, and BCB biomass source; and eight 3D variables including the velocity field in all three directions, pressure, temperature, potential temperature, relative humidity and precipitation.
    
    \begin{equation}
        \frac{du}{dt} + u\frac{du}{dx} + v\frac{du}{dy} + w(\omega,T,q,z)\frac{du}{dz} = \Sigma
        \label{eq:transport_model}
    \end{equation}
    
    The scope of this project focuses on the troposphere. Based on the pressure range for the troposphere (1000 to 200 mbar), the total number of levels required to cover the entire troposphere is between 30 and 40. However, considering the BCB concentration distribution across vertical levels, BCB concentrations drop significantly after the 20th level, making data preprocessing and training computationally intensive. Given that ML architectures with high inductive biases in the spatial domain generally focus on regions with high concentrations to avoid penalization through the loss function (Mean Squared Error in this case), the first 20 vertical levels (up to 600 hPa) were selected for processing. This vertical range covers nearly the entire BCB concentration distribution in the troposphere, where significant variation due to surface-level emissions is observed.

\begin{figure}[htbp]
        \centering
        \includegraphics[width=\linewidth]{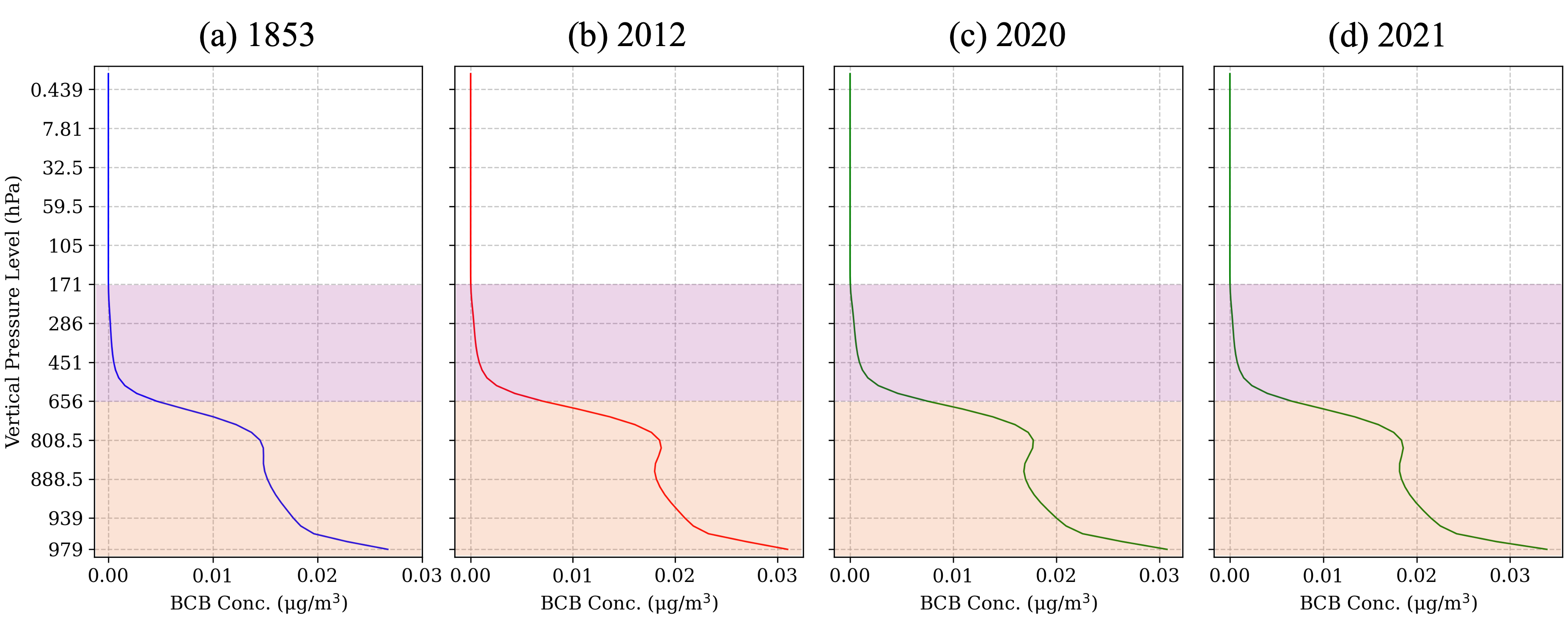}
        \caption{Vertical profiles of BCB concentration across the entire 62 atmospheric levels for four different years of the test set.}
        \label{fig:conc_vprofile_test}
    \end{figure}

\begin{figure}[htbp]
        \centering
        \includegraphics[width=\linewidth]{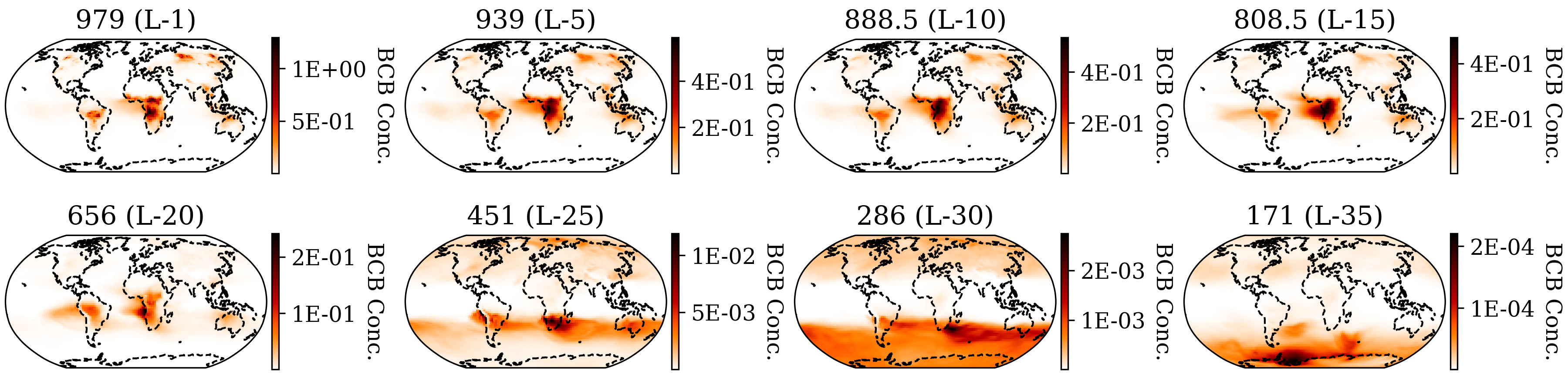}
        \caption{Spatial distribution of annual average BCB concentration during 2012 across eight different pressure levels from 979 hPa (Level 1) to 171 hPa (Level 35). Note the different color scales for each level, with concentrations decreasing by multiple orders of magnitude at higher altitudes.}
        \label{fig:conc_sdist_2012}
    \end{figure}

    Figure~\ref{fig:conc_vprofile_test} shows the BCB concentration profile across all 62 vertical levels of ModelE3 for the 4-year test period. As evident from the figure, BCB concentration begins to deplete after 808 hPa (level 15). This study estimates BCB concentration up to 656 hPa (level 20, shown in orange). BCB activities persist until 171 hPa (level 35, marked in magenta). This region falls almost entirely within the troposphere. Figure~\ref{fig:conc_sdist_2012} shows the spatial distribution of annual average BCB concentration over the test period, using 2012 as a representative example for all four years in the test set. The figure demonstrates that BCB concentration decreases by three orders of magnitude between 656 and 171 hPa (levels 20 to 35), which makes data preprocessing and training very laborious as the number of input layers and the consequent target variable output layers increase. Figure~\ref{fig:conc_sdist_2012} also shows that the global pattern of BCB concentration changes drastically after level 20.
    
\section{Model Architecture} \label{sec:appendix-2}
\begin{figure}[htbp]
        \centering
        \includegraphics[width=0.7\linewidth]{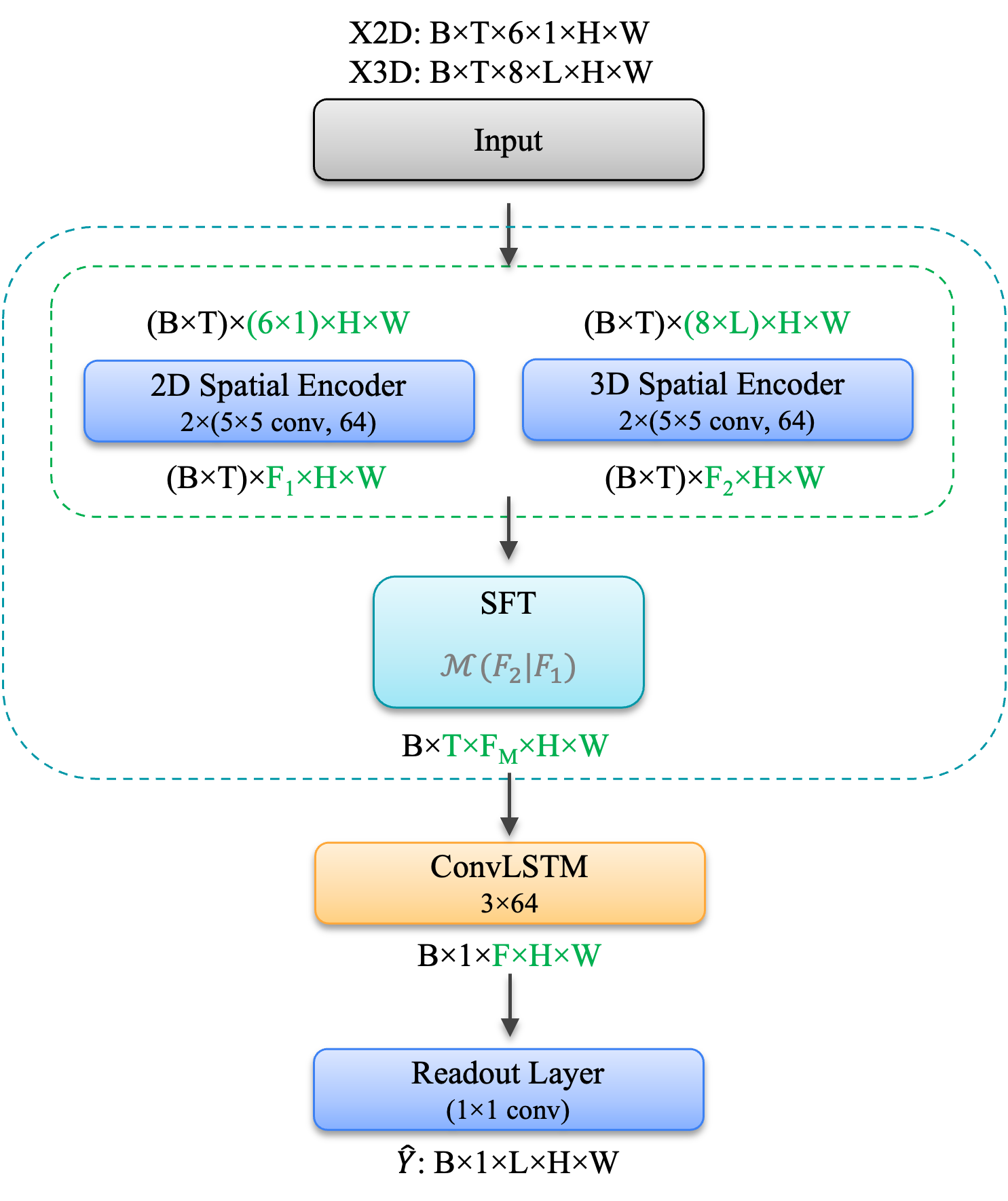}
        \caption{Spatiotemporal ML model, equipped with spatial feature extraction and modality fusion.}
        \label{fig:model_architecture}
    \end{figure}

    The model receives 2D and 3D data separately. According to Figure~\ref{fig:model_architecture}, we used spatial encoders to process two different modalities (2D and 3D variables). Suppose a batch of 3D input tensors $\mathbf{\beta} \in \mathbb{R}^{B \times T \times V \times L \times H \times W}$, where the batch size is $B = |\mathbf{\beta}|$, $V$ represents the number of 3D variables, and $L$ represents the number of associated levels for each 3D variable. In the spatial encoder, we reshape the sequential input data from $B \times T \times V \times L \times H \times W$ to $(B \times T) \times (V \times L) \times H \times W$ so that only spatial correlations are taken into account. The same procedure is applied to 2D variables with the number of levels equal to 1. The outputs of the 3D and 2D encoders are $(B \times T) \times F_1 \times H \times W$ and $(B \times T) \times F_2 \times H \times W$, respectively, where $F_1 = F_2$ is the identical number of channels in the latent space. The outputs of the encoders are fused via Spatial Feature Transform (SFT). In fact, SFT modulates 3D features conditioned on 2D features. In other words, the SFT layer generates affine transformation parameters for spatial-wise feature modulation. SFT layers can be trained end-to-end together with the network using the same loss function. This module is designed for spatial feature extraction and modality fusion. The output of this step is in the shape of $B \times T \times F_m \times H \times W$, which can be fed into any spatio-temporal ML model to impose the effect of time evolution on the latent features and map them to the target variable.

    \subsection{Experimental Settings}
        The training consists of two tiers. First, the model is trained on data from 1851--1852, comprising 35,040 timesteps, for 50 epochs. The training uses the Adam optimizer with default beta values of 0.9 and 0.999. The loss function combined Mean Squared Error (MSE) and Mean Absolute Error (MAE) with equal weights to balance sensitivity to outliers and overall error magnitude. The initial learning rate of 0.0005 and a Cosine Annealing scheduler is considered for training. To avoid unstable training, the first 10 steps are considered as a warm-up period with a factor of 0.1 and Linear scheduler. In the second tier, the model is then fine-tuned on data from 2010--2011 for 30 epochs using the Adam optimizer with an initial learning rate of $1/5$ of the first tier (i.e., 0.0001) and a Cosine Annealing scheduler. For each round of training, ten percent of the training period is randomly selected as the validation set. However, the entire years of 1853, 2012, 2020, and 2021 are reserved for testing. Therefore, the model is trained on 4 years of data and tested on 4 years, where 2 test years (2020 and 2021) represent entirely different time periods outside the training domain. This configuration provides a crucial capability for climate modeling applications requiring reliable long-term projections, as it evaluates the model's ability to generalize to future climate conditions not seen during training.

\begin{figure}[htbp]
        \centering
        \includegraphics[width=\linewidth]{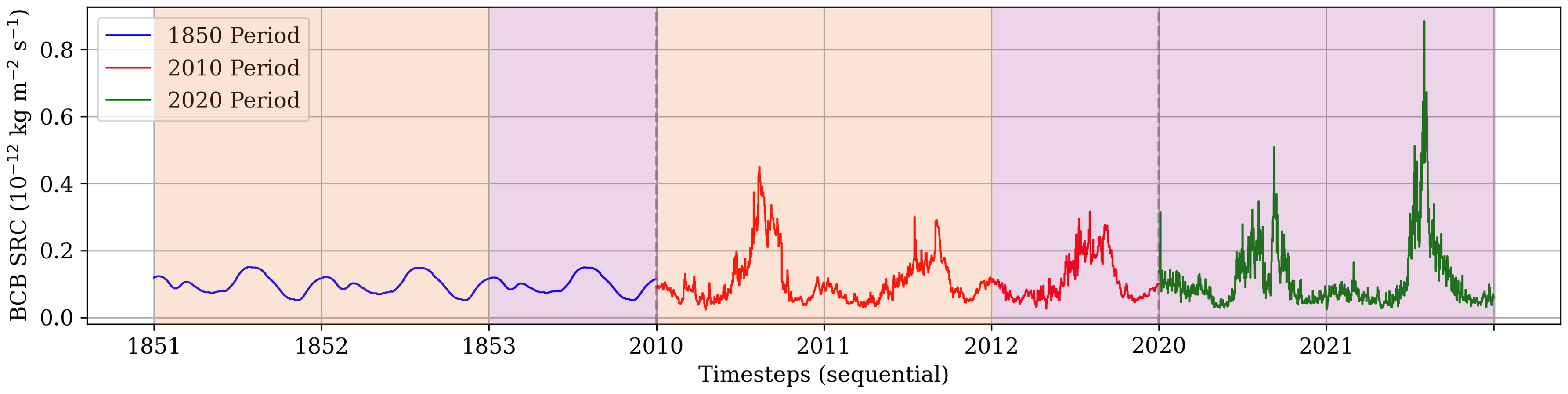}
        \caption{Global average daily BCB emission sources for pre-industrial (blue line, 1851-1853) and industrial periods (red line for 2010-2012, green line for 2020-2021). Training years are highlighted with orange shading and test years with purple shading.}
        \label{fig:train_test_years}
    \end{figure}

    Figure~\ref{fig:train_test_years} shows the global average of daily BCB emission sources for training and test years. The training years (1851-1852, 2010-2011) are marked by orange shading while the test years (1853, 2012, and 2020-2021) are highlighted by purple shading. As shown, the concentration of BCB emission sources is considerably higher during industrial periods compared to pre-industrial periods (1851 to 1853). In pre-industrial periods (blue line), the values oscillate smoothly as these daily values are calculated by interpolation of monthly averages. The industrial periods show much higher variability, with the 2020-2021 period (green line) exhibiting the most extreme peaks. Notably, the 2021 peaks reach approximately two times higher than the highest values observed in the earlier industrial period (2010-2012, red line).

\section{Results} \label{sec:appendix-3}
    Throughout this paper, we refer to atmospheric levels using both pressure values (in hPa) and level nomenclature interchangeably. Table~\ref{tab:pressure_levels} provides the complete mapping between the 20 atmospheric levels and their corresponding pressure values used in our analysis. Readers can reference this table to identify the specific atmospheric level when pressure values are mentioned in the text, figures, or discussions of model performance across different vertical layers of the atmosphere.

    \begin{table}[htbp]
        \centering
        \caption{Atmospheric pressure levels and their corresponding pressure values (hPa) used in this study.}
        \label{tab:pressure_levels}
        \begin{tabular}{llllllll}
        \hline
        Level & Press & Level & Press & Level & Press & Level & Press \\
        \hline
        level-01 & 979.0 & level-06 & 929.0 & level-11 & 877.0 & level-16 & 784.0 \\
        level-02 & 969.0 & level-07 & 919.0 & level-12 & 864.0 & level-17 & 756.5 \\
        level-03 & 959.0 & level-08 & 909.0 & level-13 & 848.5 & level-18 & 726.0 \\
        level-04 & 949.0 & level-09 & 899.0 & level-14 & 830.0 & level-19 & 692.5 \\
        level-05 & 939.0 & level-10 & 888.5 & level-15 & 808.5 & level-20 & 656.0 \\
        \hline
        \end{tabular}
    \end{table}

    \subsection{Global Distribution of Coefficient of Determination}
\begin{figure}[htbp]
        \centering
        \includegraphics[width=1\linewidth]{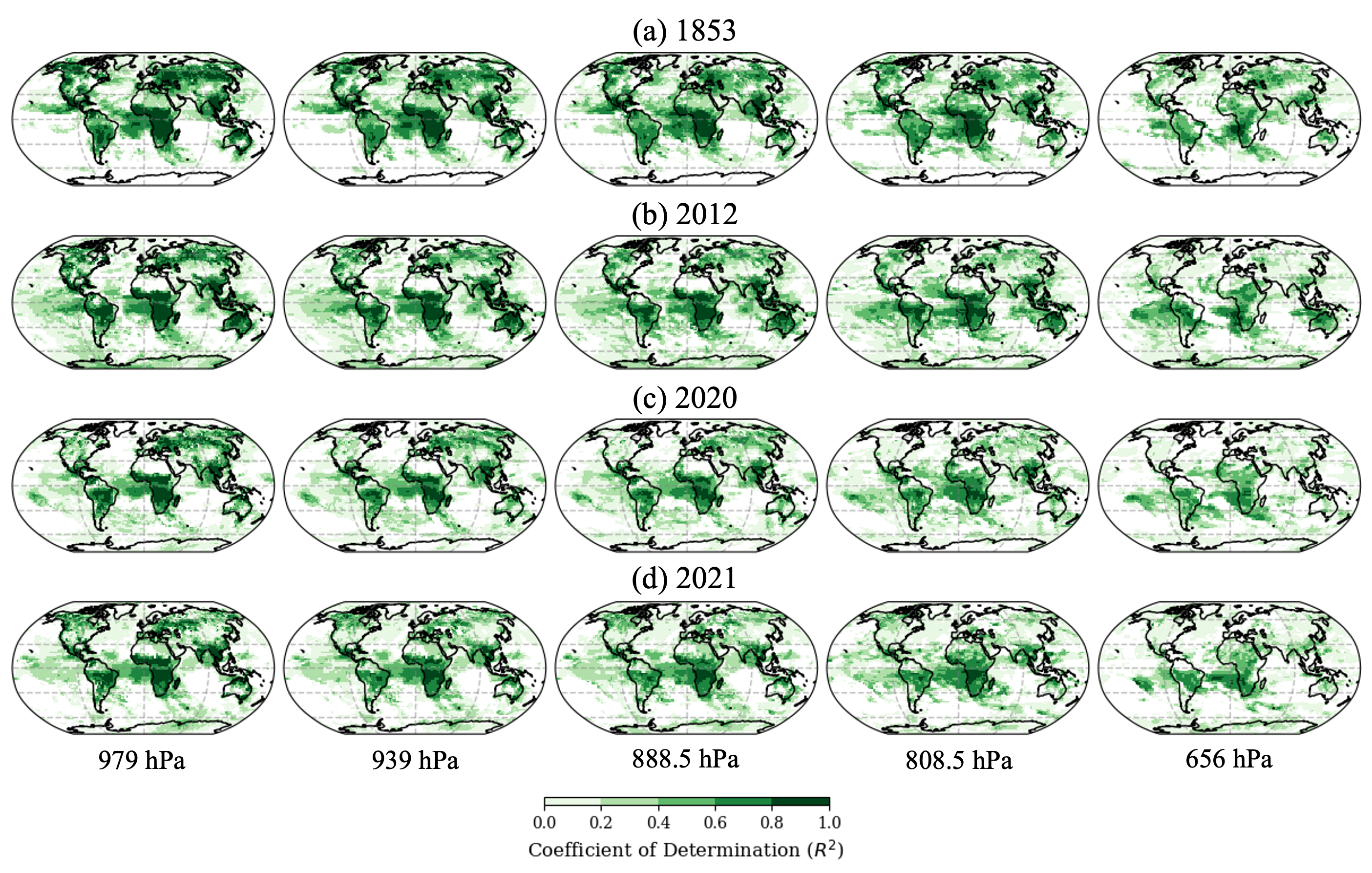}
        \caption{Global distribution of R$^2$ across different pressure levels.}
        \label{fig:rsquare}
    \end{figure}

    Figure~\ref{fig:rsquare} shows spatial variations of R$^2$ from surface pressure level (Level 01) up to Level 20, with darker green areas indicating higher R$^2$ values and white areas representing zero or negative R$^2$ values where the climatological average could perform better than the ML model in those regions. Investigation on the low performance regions shows that the BCB concentrations in these regions (with negative R$^2$ values) are smaller than those in high-concentration regions by several orders of magnitude; thus, the models generally produce random noise in these areas. 
    Across all test periods, higher R$^2$ values are observed in tropical regions (such as the Amazon and central Africa), as well as parts of the northern mid-latitude zone and Southeast Asia, indicating strong agreement between predictions and ground truth in these areas. Conversely, lower R$^2$ values are prevalent over oceanic regions and certain arid land areas, where BCB concentrations are lower in value compared to regions with major emission sources. 
    
    \subsection{BCB Global Average Over Time}
\begin{figure}[htbp]
        \centering
        \includegraphics[width=\linewidth]{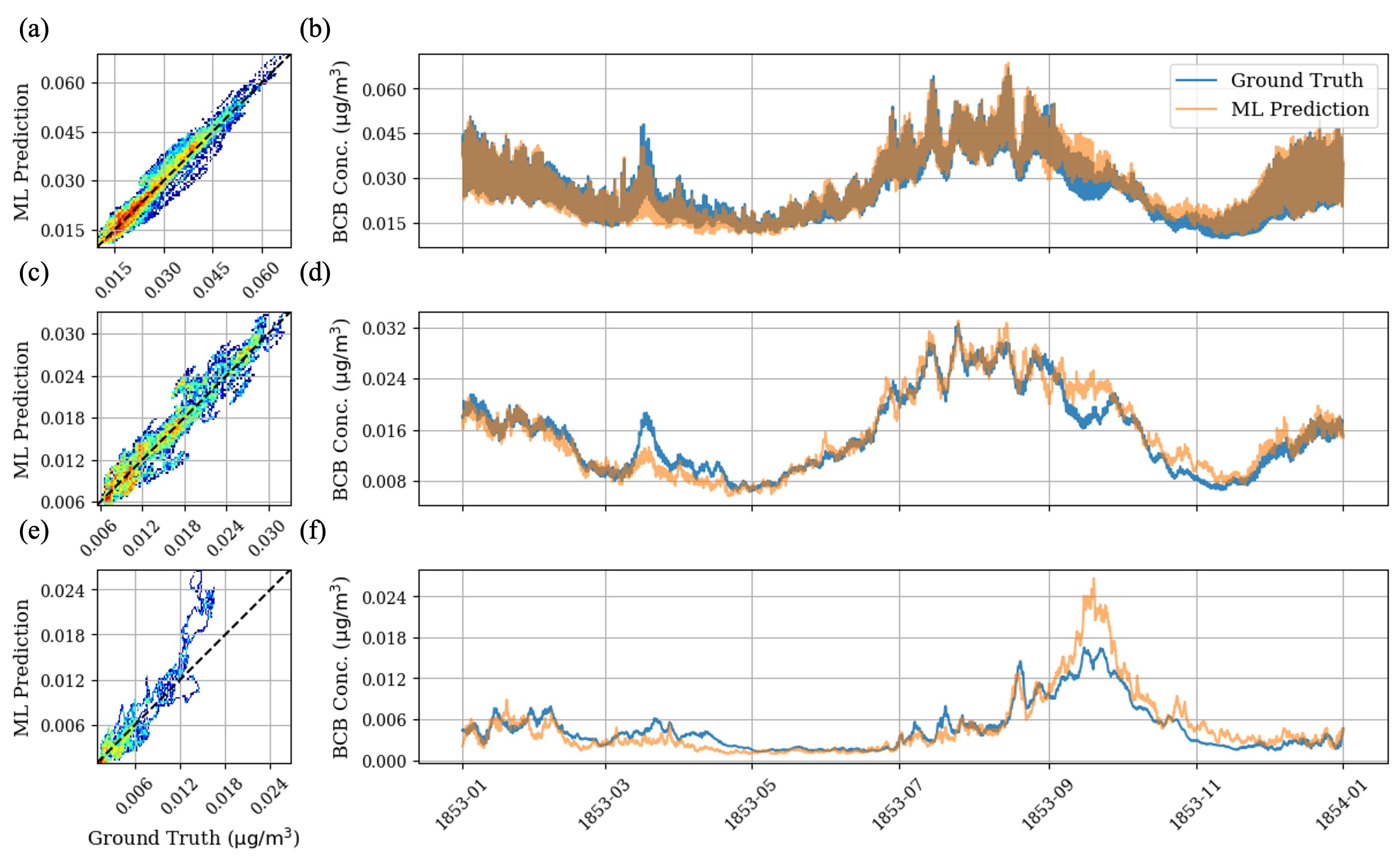}
        \caption{(a, c, e) Scatter plot comparing the global averages of model results with ground truth values for 1853. (b, d, f) Comparison of global average values of model outputs and ground truth over 1853.}
        \label{fig:global_average_1853}
    \end{figure}

\begin{figure}[htbp]
        \centering
        \includegraphics[width=\linewidth]{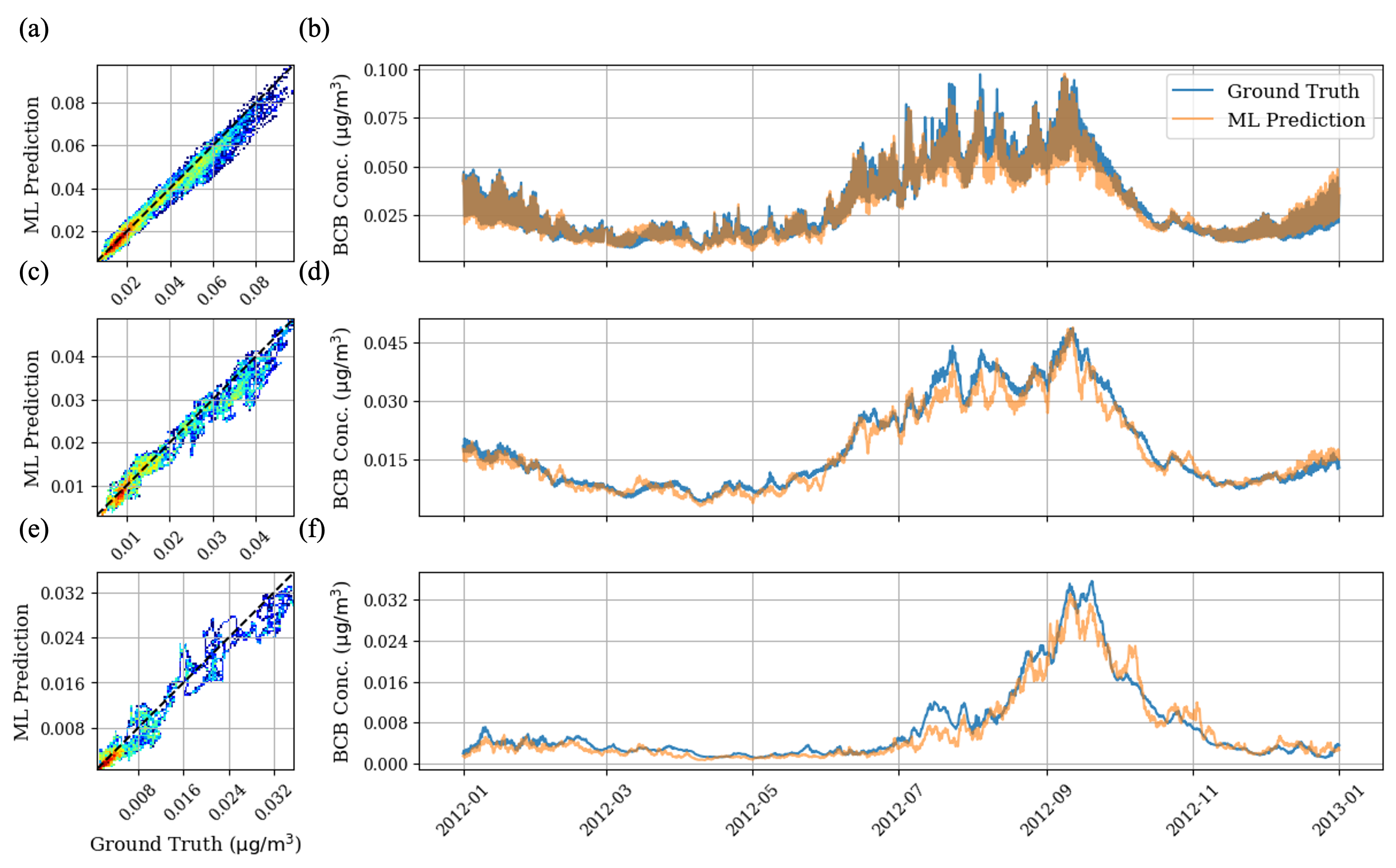}
        \caption{(a, c, e) Scatter plot comparing the global averages of model results with ground truth values for 2012. (b, d, f) Comparison of global average values of model outputs and ground truth over 2012.}
        \label{fig:global_average_2012}
    \end{figure}

\begin{figure}[htbp]
        \centering
        \includegraphics[width=\linewidth]{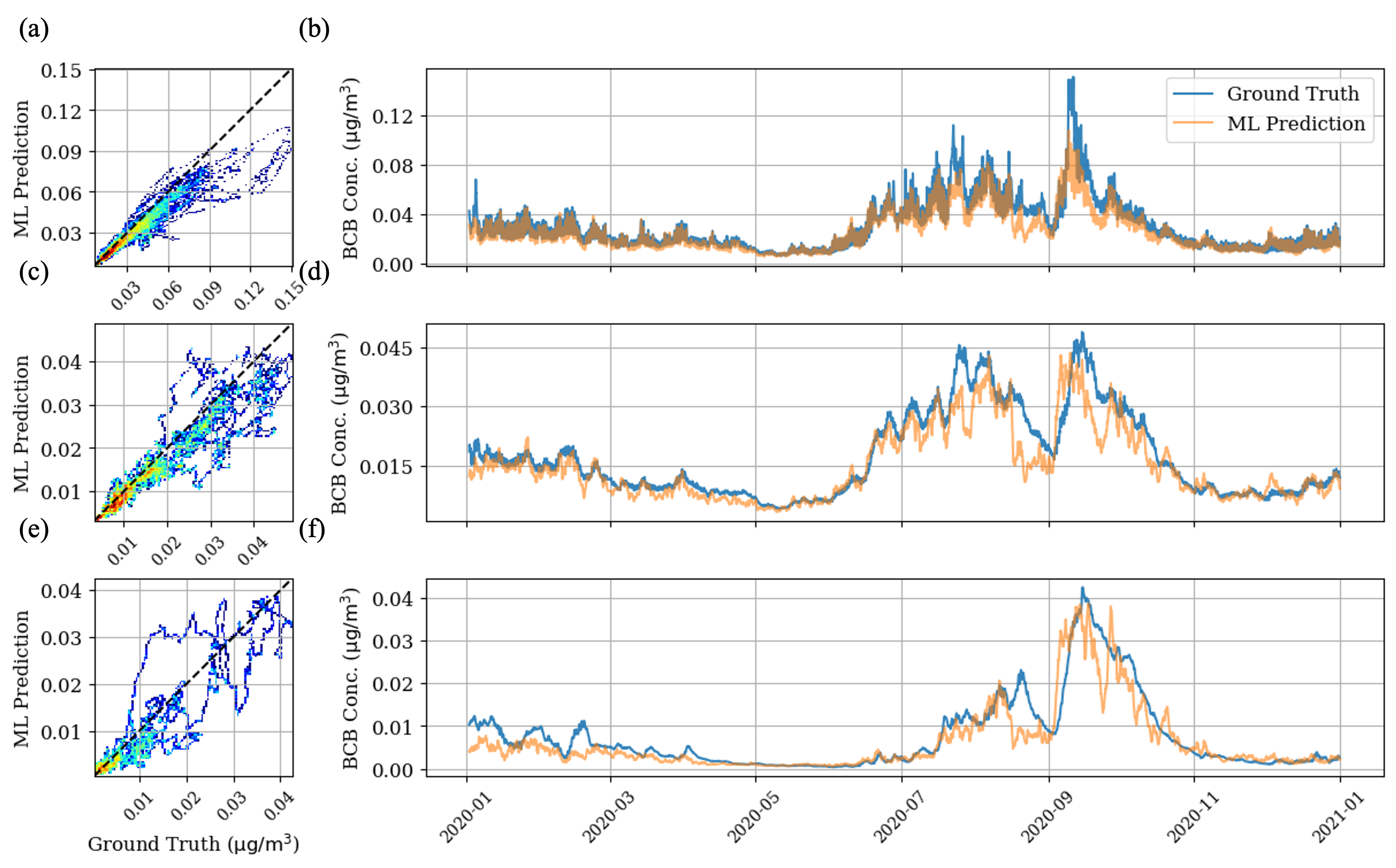}
        \caption{(a, c, e) Scatter plot comparing the global averages of model results with ground truth values for 2020. (b, d, f) Comparison of global average values of model outputs and ground truth over 2020.}
        \label{fig:global_average_2020}
    \end{figure}

\begin{figure}[htbp]
        \centering
        \includegraphics[width=\linewidth]{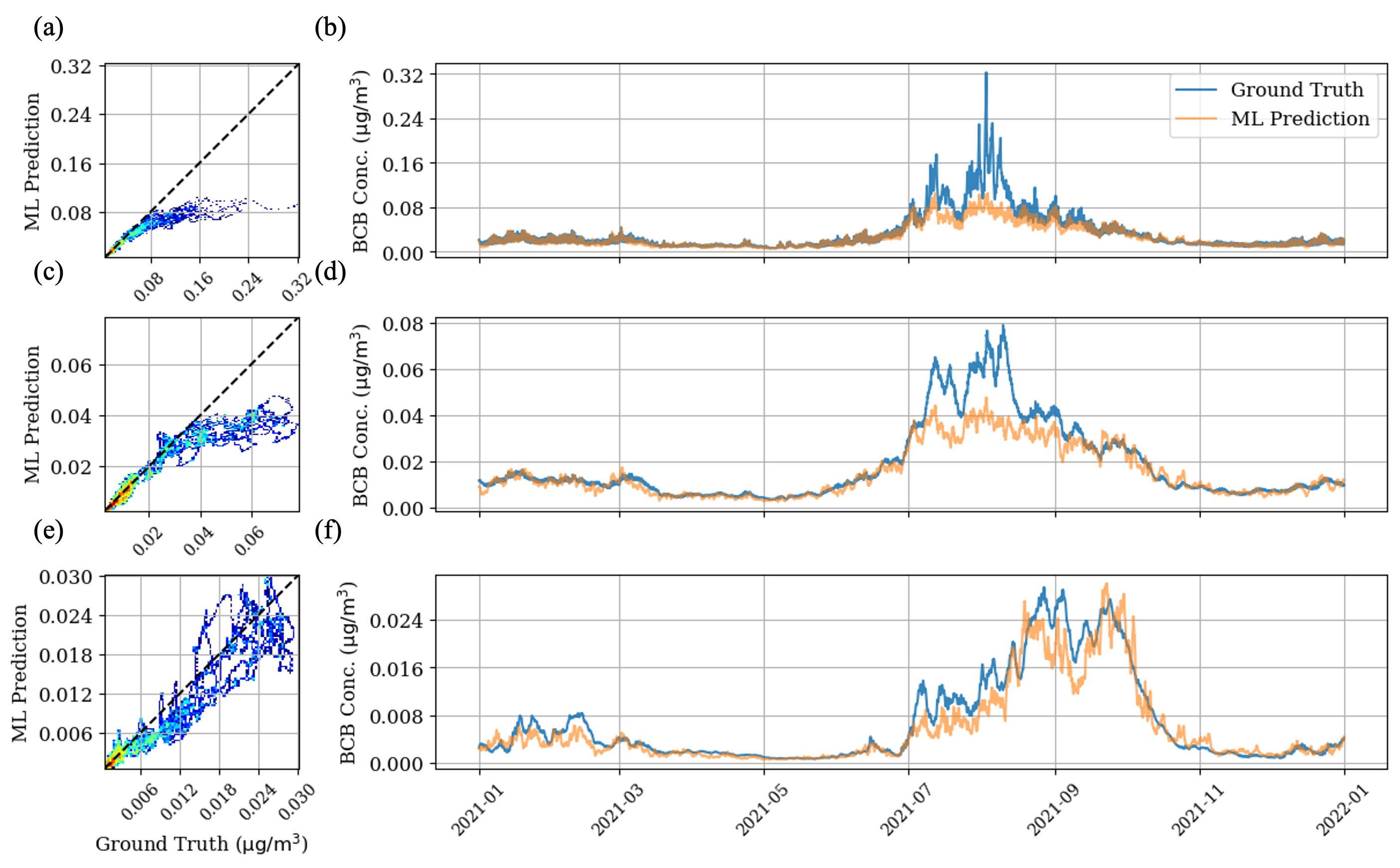}
        \caption{(a, c, e) Scatter plot comparing the global averages of model results with ground truth values for 2021. (b, d, f) Comparison of global average values of model outputs and ground truth over 2021.}
        \label{fig:global_average_2021}
    \end{figure}

    Figures~\ref{fig:global_average_1853},~\ref{fig:global_average_2012},~\ref{fig:global_average_2020}, and~\ref{fig:global_average_2021} compare the BCB global averages predicted by ML model with the corresponding ground truth values for 1853, 2012, 2020, and 2021, respectively. This comparison is conducted across three distinct pressure levels: level 1 (979 hPa), level 10 (888.5 hPa), and level 20 (656 hPa).
    
    Figures~\ref{fig:global_average_1853}-a,c,e present scatter plots of these comparisons for 1853, demonstrating that the model performs very well for the first and 10th pressure levels, although it tends to overestimate extreme values compared to the ground truth in the higher levels (level 20). This overestimation can be attributed to the model's fine-tuning on 2010 and 2011 data, years characterized by higher BCB emissions and concentrations compared to the preindustrial era (1850s). Consequently, the model has been calibrated for higher values at upper atmospheric levels. Figures~\ref{fig:global_average_1853}-b,d,f compare the global averages of the model outputs and ground truth values throughout the test period. These figures illustrate the model's ability to capture peak season values, though an overestimation at the 20th level occurs from mid-June through October.
    
    Figure~\ref{fig:global_average_2012} displays the same analysis for 2012. Results indicate that the model performs exceptionally well across all pressure levels without systematic over- or underestimation. This accuracy likely stems from the model's final training phase being fine-tuned on 2010 and 2011 data, making it particularly familiar with the BCB emission patterns and associated forcing during this time period. The results for 2020 (Figure~\ref{fig:global_average_2020}) are nearly as accurate as those for 2012, though there is some underestimation of extreme values at the first pressure level, while biases at higher levels appear random rather than systematic.
    
    Figure~\ref{fig:global_average_2021} presents the analysis for 2021, a year in which BCB concentration at the first pressure level reached unprecedented heights, approximately three times higher than those observed in 2012 and 2020 during peak periods (July through October). This exceptional increase manifests as model underestimation of extreme events. However, this trend diminishes at higher atmospheric levels and is completely resolved by the 20th level, where concentrations follow patterns more consistent with previous years, allowing the model to accurately capture high peaks. This observation highlights a significant achievement of our model architecture: as evidenced across all examples, patterns at higher atmospheric levels function independently from the first pressure level, where high concentrations of BCB (our target variable) accumulate. We attribute this successful vertical representation to the appropriate architectural provisions in the model structure that account for vertical convection and other mechanisms governing BCB transport across the atmospheric vertical profile.

\section{Ablation Study}
    We conducted comprehensive ablation studies to evaluate the contributions of individual components in our proposed network architecture. Specifically, we examined the effects of incorporating spatial feature encoders (Encoder), spatial feature transformation for modality fusion (SFT), or removing one or both of these modules. To optimize training resources, each model variant was trained on the entire year of 1852 and evaluated on 2012 data. Table~\ref{tab:ablation_study} presents the performance metrics for different model variants predicting the column burden of BCB, calculated by summing the concentration across all 20 pressure levels for each grid cell. Our results demonstrate that the benchmark model with all components achieves superior performance compared to reduced variants. Removing the SFT module caused a modest decrease in performance, while removing the Encoder module resulted in more significant degradation. Interestingly, the model without both modules slightly outperformed the variant missing only the Encoder, suggesting complex interactions between these architectural components. These findings highlight the critical role of both spatial feature extraction and modality fusion in achieving optimal predictive performance.

    \begin{table}[htbp]
        \centering
        \caption{Ablation study of each proposed component on performance metrics.}
        \label{tab:ablation_study}
        \begin{tabular}{lcccc}
        \hline
        \textbf{Encoder} & \textbf{SFT} & \textbf{R$^2$ $\uparrow$} & \textbf{MAE $\downarrow$} & \textbf{RMSE $\downarrow$} \\
        \hline
         &  & 0.9439 & 0.0119 & 0.0235 \\
        \checkmark &  & 0.9424 & 0.0123 & 0.0242 \\
         & \checkmark & 0.9492 & 0.0117 & 0.0227 \\
        \checkmark & \checkmark & \textbf{0.9585} & \textbf{0.0104} & \textbf{0.0199} \\
        \hline
        \end{tabular}
    \end{table}

\begin{figure}[htbp]
        \centering
        \includegraphics[width=0.5\linewidth]{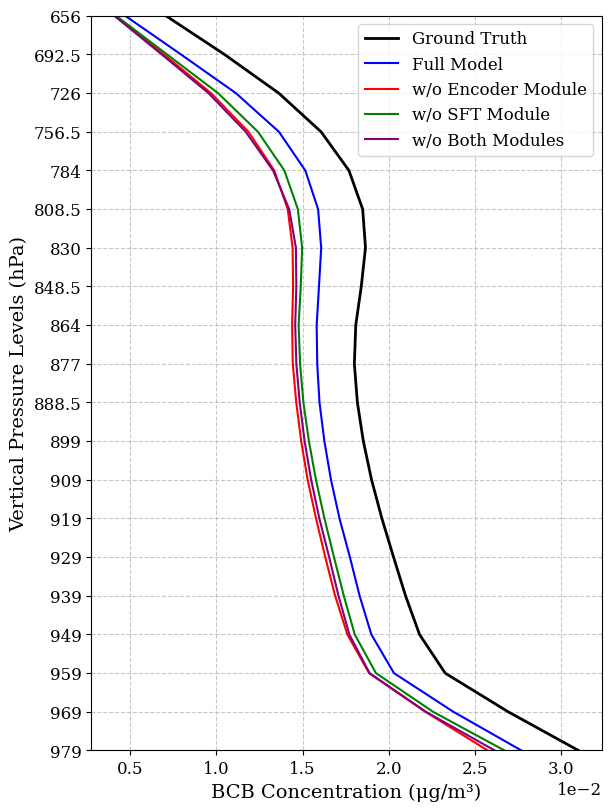}
        \caption{Impact of model components on vertical profiles of BCB concentration across 20 atmospheric levels.}
        \label{fig:ablation_study}
    \end{figure}

    Figure~\ref{fig:ablation_study} compares the vertical profiles of BCB concentration across 20 atmospheric pressure levels for different model variants. The results demonstrate that the full model most accurately follows the ground truth vertical distribution, with notably lower bias compared to variants lacking spatial feature encoders, spatial feature transformation modules, or both. This visual comparison complements the quantitative metrics in Table~\ref{tab:ablation_study}, illustrating that the full model's superior performance extends beyond integrated column burden to accurately representing concentrations at individual pressure levels throughout the atmospheric profile.
    
\end{document}